\documentstyle[13lomcon,cite,epsfig,amssymb,xcolor,ulem]{article}

\begin{document}

\title{ELECTRON ANGULAR CORRELATION IN NEUTRINOLESS DOUBLE BETA DECAY AND NEW PHYSICS}

\author{ A.~Ali \footnote{e-mail: ahmed.ali@desy.de}}

\address{ Deutsches Elektronen-Synchrotron, DESY, 22607 Hamburg,
Germany}

\author{ A.V.~Borisov \footnote{e-mail: borisov@phys.msu.ru}, D.V.~Zhuridov \footnote{e-mail:
dmitry.zhuridov@uwr.edu.pl}}

\address{Faculty of Physics, Moscow State University, 119991 Moscow,
Russia}

\maketitle\abstracts{The angular correlation of the electrons in
the neutrinoless double beta decay ($0\nu2\beta$) is calculated
taking into account the nucleon recoil, the $S$ and $P$-waves for
the electrons and the electron mass using a general Lorentz
invariant effective Lagrangian. We show that the angular
coefficient is essentially independent of the nuclear matrix
element models. We work out the angular coefficient in several
scenarios for new physics, in particular, in the left-right
symmetric models.}

\section{Introduction}
It is now established that the observed neutrinos have tiny masses
and they mix with each other~\cite{PDG}. Theoretically, it is
largely anticipated that the neutrinos are Majorana particles.
Experimental evidence for $0\nu2\beta$ decay would deliver a
conclusive confirmation of the Majorana nature of neutrinos,
establishing the existence of physics beyond the
SM~\cite{Vogel:2006sq}. An extended version of the SM could
contain tiny nonrenormalizable terms that violate lepton number
(LN) and allow the $0\nu2\beta$ decay. Probable mechanisms of LN
violation may include exchanges by: Majorana neutrinos $\nu_M$s
~\cite{ZKS,Doi} (the preferred mechanism after the observation of
neutrino oscillations \cite{PDG}), SUSY Majorana
particles~\cite{SUSY1,SUSY}, scalar bilinears (SBs)~\cite{BL},
e.g. doubly charged dileptons (the component of the $SU(2)_L$
triplet Higgs etc.), leptoquarks (LQs)~\cite{LQ}, right-handed
$W_R$ bosons~\cite{Doi,HKP} etc. From these particles light $\nu$s
are much lighter than the electron and others are much heavier
than the proton that gives two possible classes of mechanisms for
the $0\nu2\beta$ decay: long range (with the light $\nu$s in the
intermediate state) and short range mechanism. Our aim was to
examine the possibility to discriminate among the various possible
mechanisms contributing to the $0\nu2\beta$-decays and the various
sources of LN violation using the information on the angular
correlation of the final electrons. We published a preliminary
study along these lines in Ref.~\cite{Ali:2006iu} and a more
detailed study in Ref.~\cite{ABZ_PRD}. Here, we summarize the main results of Ref. ~\cite{ABZ_PRD}.

\section{Angular correlation for the long range mechanism of $0\nu2\beta$ decay}
\label{section_Lagrangian} For the decay mediated by light
$\nu_M$s, the most general effective Lagrangian is the Lorentz
invariant combination of the leptonic $j_\alpha$ and the hadronic
$J_\alpha$ currents of definite tensor structure and chirality
\cite{Limits,Gamov}
\begin{equation}\label{L}
    {\mathcal L}=\frac{G_FV_{ud}}{\sqrt{2}}[(U_{ei}+\epsilon^{
    V-A}_{V-A,i})j_{V-A}^{\mu
    i}J^+_{V-A,\mu}+\sum\limits_{\alpha,\beta}\!^{^\prime}
\epsilon^\beta_{\alpha i}j^i_\beta J^+_\alpha+{\rm h.c.}]~,
\end{equation}
where the hadronic and leptonic currents are defined as:
$J^+_\alpha=\bar{u} O_\alpha d$ and $j^i_\beta =\bar{e} O_\beta
\nu_i$; the leptonic currents contain neutrino mass eigenstates,
and the index $i$ runs over the light eigenstates. Here and
thereafter, a summation over the repeated indices is assumed;
$\alpha$,\,$\beta$=$V\!\mp\!A$,\,$S\!\mp\!P$,\,$T_{L,R}$
($O_{T_\rho}=2\sigma^{\mu\nu}P_\rho$,
$\sigma^{\mu\nu}=\frac{i}{2}\left[\gamma^\mu,\gamma^\nu\right]$,
$P_\rho=(1\mp \gamma_5)/2$ is the projector, $\rho=L,\,R$); the
prime indicates the summation over all the Lorentz invariant
contributions, except for $\alpha=\beta=V-A$, $U_{ei}$ is the PMNS
mixing matrix. The coefficients $\epsilon_{\alpha i}^\beta$ encode
new physics, parametrizing deviations of the Lagrangian from the
standard $V-A$ current-current form and mixing of the non-SM
neutrinos.

The nonzero $\epsilon_\alpha^\beta$ for the particular SM
extensions are collected in Table 1.
\vspace*{2mm}
\begin{center}
{Table 1.}

\begin{tabular}{|c|c|}
  \hline
   Model & Nonzero $\epsilon$s \\
  \hline
   with $W_R$s &  $\epsilon_{V\mp A}^{V\mp A}$  \\
  \hline
   RPV SUSY & $\epsilon_{S+P}^{S\mp P}$, $\epsilon_{V-A}^{V-A}$, $\epsilon_{T_R}^{T_R}$  \\
  \hline
   with LQs & $\epsilon_{S\mp P}^{S+P}$, $\epsilon_{V\mp A}^{V+A}$  \\
  \hline
\end{tabular}
\end{center}
\vspace*{2mm}

We have calculated the leading order in the Fermi constant and the
leading contribution of the parameters $\epsilon_\alpha^\beta$
using the approximation of the relativistic electrons and
nonrelativistic nucleons. We take into account the $S_{1/2}$ and
the $P_{1/2}$ waves for the outgoing electrons and include the
finite de Broglie wave length correction for the $S_{1/2}$ wave.
Taking into account the nucleon recoil terms including the terms
due to the pseudoscalar form factor we obtain the differential
width in $\cos \theta$ for the
$0^+\!(A,Z)\rightarrow\!0^+\!(A,Z+2) e^- e^-$ transitions:
\begin{eqnarray}\label{dG}
\frac{d\Gamma}{d\cos\theta} = \frac{\ln2}{2}|M_{GT}|^2{\mathcal
A}(1-K\cos\theta),
\end{eqnarray}
where $\theta$ is the angle  between the electron momenta in the
rest frame of the parent nucleus, $M_{GT}$ is the Gamow--Teller
nuclear matrix element and the angular correlation coefficient is
\begin{equation}\label{K}
K={\mathcal B}/{\mathcal A}~,\quad -1<K< 1.
\end{equation}
Expressions for $\mathcal{A}$ and $\mathcal{B}$ are given in
Ref.~\cite{ABZ_PRD}. The analytic expressions associated with the
coefficients $\epsilon_{V\mp A}^{V+A}$ confirm the results of Ref.
\cite{Doi}, while the expressions associated with $\epsilon_{V\mp
A}^{V-A}$, $\epsilon_{S\mp P}^{S\mp P}$,
$\epsilon_{T_{L,R}}^{T_{L,R}}$ transcend the earlier work.


\section{Analysis of the electron angular correlation}
Consider the case of zero effects of all the interactions beyond
the SM extended by the $\nu_M$s (i.e., all
$\epsilon^\beta_\alpha=0$), which we call the ``nonstandard"
effects.  The values of $K=K_0\equiv K(\epsilon^\beta_\alpha=0)$ for various decaying nuclei are given
in Table~2.

\vspace*{1mm}

\begin{center}
{Table 2.}

\begin{tabular}{|c|c|c|c|c|c|}
  \hline
         &$^{76}\mbox{Ge}$      & $^{82}\mbox{Se}$       &$^{100}\mbox{Mo}$&$^{130}\mbox{Te}$ &$^{136}\mbox{Xe}$ \\
  \hline
     $K$ & 0.81         &  0.88         &  0.88         &  0.85        &   0.84   \\
  \hline
\end{tabular}
\end{center}
\vspace*{2mm}

We will concentrate on the case of $^{76}$Ge nucleus in the
following. Using Table~1 and taking into account the fact that
$|\mu_\alpha^\beta|$ are suppressed in comparison with
$|\epsilon_\alpha^\beta|$ by the factor $m_i/m_e$ (the chiral
suppression), we find the coefficient $K$ and the set
$\{\epsilon\}$ of nonzero $\epsilon_\alpha^\beta$s that can change the
$1-0.81\cos\theta$ form of the correlation for the SM plus
$\nu_M$s, see Table 3 (the lower three entries).
\vspace*{2mm}
\begin{center}
{Table 3.}

\begin{tabular}{|c|c|c|}
  \hline
  SM extension & $\{\epsilon\}$ & $K$ \\
  \hline
  $\nu_M$ & --- & 0.81 \\
  \hline
  $\nu_M$+RPV SUSY & $\epsilon_{S+P}^{S+P}$, $\epsilon_{T_R}^{T_R}$ & $-1<K<1$ \\
  \hline
  $\nu_M+$RC & $\epsilon_{V\mp A}^{V+A}$ & $-1<K<1$ \\
  \hline
  $\nu_M+{\rm SLQ}$ & $\epsilon_{S\mp P}^{S+P}$ & $-1<K<1$ \\
  \hline
\end{tabular}
\end{center}
\vspace*{2mm}
They correspond to the following extensions of the SM:  $\nu_M$s
plus RPV SUSY \cite{SUSY}, $\nu_M$s plus right-handed currents
(RC) (connected with right-handed $W$ bosons \cite{Doi} or vector LQs
\cite{LQ}), and $\nu_M$s plus scalar LQs~\cite{LQ}. Hence, $K$ can signal the presence of this new
physics.

Let us now consider some particular cases for the parameter space.
We will analyze only the terms with $\epsilon_{V\mp A}^{V\mp A}$
as the corresponding nuclear matrix elements have been worked out
in the literature. We use various types of
QRPA~\cite{Pantis,Kortelainen:2007rh}.

In the case of $|\langle m\rangle|=0$ the current lower bound
$T_{1/2}>1.6\times 10^{25}~\mbox{yr}$ for the $^{76}\mbox{Ge}$
nucleus~\cite{Aalseth} yields the upper bounds on the parameters
$|\mu_{V\mp A}^{V-A}|$, $|\epsilon_{V\mp A}^{V+A}|$ that give
bounds on the parameters of the particular models~\cite{ABZ_PRD}.
The fact that the dependence of $K$ on the nuclear matrix elements
is much weaker than the uncertainty in $T_{1/2}$ from this source
was illustrated in Ref. \cite{ABZ_PRD}.

In the case of $|\langle m\rangle|\neq 0$, $\cos\psi_i=0$, where
$i$ depends on $\alpha$, $\beta$, for $\epsilon_{V+A}^{V+A}\neq 0$
\begin{eqnarray}\label{Eq:V+A}
  |\mu|^2 = (7.9+10K)\times10^{12}/T_{1/2}, \quad
    |\epsilon_{V+A}^{V+A}|^2 = (5.1-6.3K)\times10^{12}/T_{1/2},
\end{eqnarray}
with $T_{1/2}$ in years, and for $\epsilon_{V-A}^{V+A}\neq 0$
\begin{eqnarray}\label{Eq:V-A}
  |\mu|^2 = (7.7+10K)\times10^{12}/T_{1/2}, \quad
    |\epsilon_{V-A}^{V+A}|^2 = (1.9-2.4K)\times10^{8}/T_{1/2}.
\end{eqnarray}
The correlations among $|\epsilon_{V\mp A}^{V+A}|$, $T_{1/2}$, $K$
were used in the analysis of left-right symmetric
models~\cite{LR}. In the model $SU(2)_L\times SU(2)_R\times U(1)$
we have
\begin{equation}
\label{WR} m_{W_R} =
m_{W_L}\left(\epsilon/\left|\epsilon^{V+A}_{V+A}\right|\right)^{1/2},
\quad
\zeta=-\arctan\left(\left|\epsilon^{V+A}_{V-A}\right|/\epsilon\right)
\end{equation}
for the mass of the right-handed $W$ boson and its mixing angle
$\zeta$ with the left-handed one. The correlation among $m_{W_R}$
($\zeta$), $K$, and $T_{1/2}$ is shown in Fig. 1~{\it left} ({\it
right}) for conservative value $\epsilon=10^{-6}$ for the mixing
parameter $\epsilon=|U_{ei}V_{ei}|$. It is clear that the closer
is $K$ to 1 for the fixed value of $T_{1/2}$ the stronger is the
lower bound on $m_{W_R}$ (the upper bound on $\zeta$).
\begin{figure}
\vspace{-1.6cm} \hspace{-1.5cm}
\begin{minipage}{0.49\textwidth}
\includegraphics[scale=0.7]{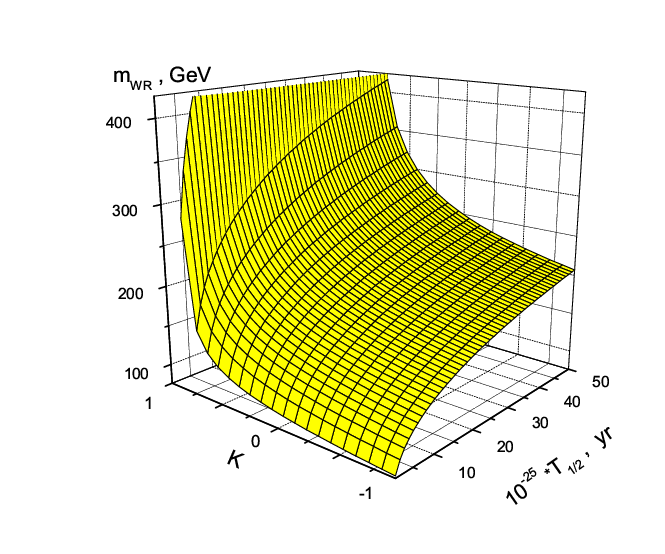}
\end{minipage}
\begin{minipage}{0.49\textwidth}
\includegraphics[scale=0.7]{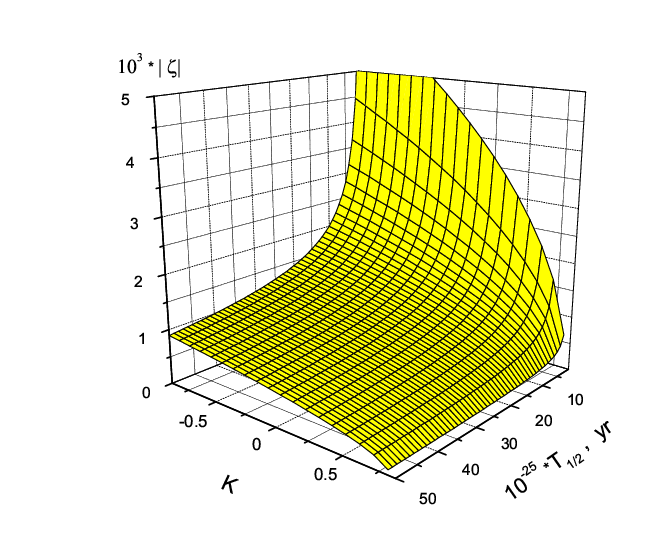}
\end{minipage}
\caption{Correlation between the right-handed
$W$-boson mass $m_{W_R}$ ({\it left}) or the mixing $\zeta$ ({\it
right}), the angular coefficient, and the half-life $T_{1/2}$ for
the $0\nu2\beta$ decay of $^{76}\mbox{Ge}$.}
\end{figure}
We have shown also that the sensitivity of the angular correlation
to the $W_R$ mass increases with decreasing values of the
effective Majorana neutrino mass $|\langle m\rangle|$
\cite{ABZ_PRD}.

\section*{Acknowledgments}
We thank Alexander Barabash and Fedor \v{S}imkovic for helpful
discussions. One of us (DVZ) would like to thank DESY for the
hospitality in Hamburg where a good part of this work was done.

\section*{References}

\end{document}